\documentclass[12pt]{article}
\usepackage{amsmath}
\usepackage{bm}

\begin{document}
\title{Entangled states in supersymmetric quantum mechanics}

\author{
H. P. Laba$^1$,
V. M. Tkachuk$^2$\footnote{voltkachuk@gmail.com}\\
$^1$Department of Applied Physics and Nanomaterials Science, \\
Lviv Polytechnic National University,\\
5 Ustiyanovych St., 79013 Lviv, Ukraine,\\
$^2$Department for Theoretical Physics,\\
Ivan Franko National University of Lviv,\\
12, Drahomanov St., Lviv, 79005, Ukraine.}

\maketitle

\begin{abstract}

We study entanglement of spin degrees of freedom with continuous one in supersymmetric (SUSY) quantum mechanics.
Concurrence is determined by mean value of spin and is calculated explicitly for SUSY states.
We show that eigenstates of supercharges are maximally entangled. As an example
the entanglement of atom state with photon state and SUSY in Jaynes-Cummings model are considered.


\end{abstract}

\section{Introduction}

In 1981 Witten introduced supersymmetric (SUSY) quantum mechanics as a
laboratory for investigating main features of supersymmetry \cite{Wit81}.
Some time before Nicolai showed that SUSY could also be a useful tool for studies of spin
systems \cite{Nic76}. Subsequently it was recognized that SUSY quantum mechanics is
interesting on its own right. For review of SUSY quantum mechanics see for instance \cite{Hau86,Coo95,Gan17}.

It was shown that SUSY quantum mechanics is powerful tool
for searching exact solutions of eigenvalue problems in quantum mechanics (see, for instance, \cite{Gen83,Tka99,Kul99,Que03,Que18} and references therein).
It was also shown that there are different quantum problems where SUSY is presented as physical symmetry.
For instance, it was found
that the motion of electron in
magnetic field possess SUSY ( see for instance \cite{Tka97});
the  equations describing propagation of electromagnetic waves in planar waveguide can be casted in SUSY form \cite{Lab14}(see also \cite{Lon15,Mac18});
some models in quantum optics possess SUSY \cite{Tom15}.

Quantum entanglement is
the intriguing feature of quantum theory and has been noted by
Einstein, Podolsky, Rosen, and Shr$\ddot{\rm o}$dinger \cite{Ein35,Schrod35}
(see for review \cite{Hor09,Ben17}).
As a result it was recognized that entanglement is a new important resource for
quantum-information processing \cite{Wen17,Fla19}.

The entanglement in SUSY Bose-Fermi system was studied in \cite{Ili04}.
In \cite{TkaVak07}  entanglement of spin variables with continuous variables of electron in uniform magnetic field which exhibit
SUSY was examined.
Recently it was also noted about the entanglement of spin and continuous degrees of freedom in SUSY quantum mechanics
\cite{Cat14}.
In paper \cite{Mot18} the authors studied the entanglement, squeezing and statistical
properties of supercoherent states and calculated the concurrence for these states.
In this paper we study entanglement for quantum states  of SUSY quantum mechanics.
We show that concurrence is determined by mean value of spin and firstly find explicit expression for concurrence for SUSY states.
We also show that maximally entangled states are eigenstates of supercharges.

This paper is organized as follows. In Section 2 we present short introduction to SUSY quantum mechanics and give definitions which are necessary
for studies of quantum entanglement in SUSY states. In Section 3 we study entanglement of SUSY quantum states and calculate
the concurrence of these states. In Section 4  the entanglement of atom state with photon state and SUSY  in Jaynes-Cummings model are considered. Conclusions are presented in Section 5.

\section{SUSY quantum mechanics
%SUSY algebra
}

Supersymmetry means that Hamiltonian is invariant under transformation of boson into fermion $Q^+=gbf^+$ and vice versa $Q^-=(Q^+)^+=g^*b^+f$,
where $b, b^+$ and $f,f^+$ are Bose and Fermi annihilation and creation operators.
Bose and Fermi operators of creation and annihilation satisfy
the well known permutation relations
\begin{eqnarray}
[b, b^+]=1, \ \ \{f,f^+\}=1, \ \ f^2=(f^+)^2=0
\end{eqnarray}
and Bose operators commute with Fermi ones.
Taking into account that $f^2=(f^+)^2=0$ we find
$(Q^{\pm})^2=0$. Then the simplest Hamiltonian, commuting with $Q^{\pm}$, is $H=\{Q^+,Q^-\}$.
In such a way we obtain $N=2$ SUSY algebra
\begin{eqnarray} \label{Q+Q-}
\{Q^+,Q^-\}=H,  \ \ (Q^{\pm})^2=0,
\ \ [Q^{\pm}, H]=0.
\end{eqnarray}
Here it is important to note  that generators of supersymmetry written in general form $Q^+=Bf^+$ and $Q^-=(Q^+)^+=B^+f^-$
(here $B=B(b,b^+)$ is a function of Bose operators of creation and annihilation, and $B^+=(B(b,b^+))^+$)
preserve the $N=2$ SUSY algebra (\ref{Q+Q-}).

It is convenient to introduce hermitian operators which are called supercharges
\begin{eqnarray}\label{Superchar}
Q_1=Q^+ +Q^-, \ \ Q_2={1\over i}(Q^+ -Q^-).
\end{eqnarray}
One can verify that these supercharges satisfy the following algebra
\begin{eqnarray}\label{QiQj}
\{Q_i,Q_j\}=2\delta_{ij}H, \ \
[Q_i,H]=0.
\end{eqnarray}
The last form of SUSY algebra means that
Hamiltonian can be written as
\begin{eqnarray}
H=Q_1^2=Q_2^2.
\end{eqnarray}

Thus the first signal that physical system can possess SUSY is that the Hamiltonian can be written
as $H=Q^2$. Then if, in addition, there is operator anticommuting with $Q$, namely $\{T,Q\}=0$, we can write two
supercharges $Q_1=Q$ and $Q_2=iTQ$ satisfying algebra (\ref{QiQj}).
Operator $T$ is called Witten's parity operator. It is worth noting that this presentation of supercharges is very usefull for searching
sypersymmetry in quantum systems.

Finally, we would like to note that $N=2$ SUSY leads to two-fold degeneracy of non-zero energy levels and it can be easily proved
using SUSY algebra.

%\section{Quantum states in SUSY quantum mechanics}

Let us consider coordinate representation for Boson operators
\begin{eqnarray}
B={1\over\sqrt2}\left({d\over dx}+W(x)\right), \ \ B^+={1\over\sqrt2}\left(-{d\over dx}+W(x)\right),
\end{eqnarray}
where $W(x)$ is superpotential and write the Fermi operators in matrix representation
\begin{eqnarray}
f=\sigma^-=\left(
             \begin{array}{cc}
               0 & 0 \\
               1 & 0 \\
             \end{array}
           \right), \ \
           f^+=\sigma^+=\left(
             \begin{array}{cc}
               0 & 1 \\
               0 & 0 \\
             \end{array}
           \right).
\end{eqnarray}
Then Hamiltonian of supersymmetric quantum mechanics has diagonal form
\begin{eqnarray}
H=\left(
    \begin{array}{cc}
      H^+ & 0 \\
      0 & H^- \\
    \end{array}
  \right),
\end{eqnarray}
where SUSY partners reads
\begin{eqnarray}
H^+=BB^+={1\over 2}\left( -{d^2\over dx}+W^2+W'\right), \\
H^-=B^+B={1\over 2}\left( -{d^2\over dx}+W^2-W'\right).
\end{eqnarray}
Thus SUSY Hamiltonian can be written in the form
\begin{eqnarray}
H= -{1\over 2}{d^2\over dx}+{1\over 2}W^2+\sigma_z W'.
\end{eqnarray}
This Hamiltonian describes one-dimensional motion of a particle with spin $s=1/2$ in scalar potential $W^2/2$
and in magnetic field $W'$ directed along $z$ axis. In relation with this it is worth noting that SUSY also is presented
in the case of motion of electron in magnetic field in two- and three-dimensional spaces (see, for instance, \cite{Tka97}).

Now we consider eigenvalue equation of SUSY Hamiltonian
\begin{eqnarray}
H\psi=E\psi.
\end{eqnarray}
Taking into account that Hamiltonian is diagonal and spin $\sigma_z$ commutes with it,
we can find eigenstates with spin up
\begin{eqnarray}\label{psiup}
\psi_{\uparrow}=\psi_E^+|\uparrow\rangle,
\end{eqnarray}
and spin down
\begin{eqnarray}\label{psido}
\psi_{\downarrow}=\psi_E^-|\downarrow\rangle,
\end{eqnarray}
where spin states
\begin{eqnarray}
|\uparrow\rangle=\left(
                   \begin{array}{c}
                     1 \\
                     0 \\
                   \end{array}
                 \right),\ \
|\downarrow\rangle=\left(
                     \begin{array}{c}
                       0 \\
                       1 \\
                     \end{array}
                   \right)
\end{eqnarray}
and $\psi_E^+$, $\psi_E^-$ satisfy eigenvalue equations for SUSY partners $H^+$ and $H^-$
\begin{eqnarray}\label{BB+}
BB^+\psi_E^+=E\psi_E^+, \\
B^+B\psi_E^-=E\psi_E^-, \label{B+B}
\end{eqnarray}
respectively.
Applying operator $B^+$ to the eigenvalue equation for $H^+$ (\ref{BB+})  we find
\begin{eqnarray}
B^+BB^+\psi_E^+=EB^+\psi_E^+.
\end{eqnarray}
As we see $B^+\psi_E^+$ is eigenfunction of $H^-$, e.i. $B^+\psi_E^+=c\psi_E^-$.
Similarly applying $B$ to the eigenvalue equation for $H^-$ (\ref{B+B})
we obtain $B\psi_E^-=c\psi_E^+$. From normalization condition of wave functions the normalization constant
reads $c=\sqrt E$. Thus we have the following transformation
\begin{eqnarray}\label{Eqzero}
B^+\psi_E^+=\sqrt E \psi_E^-, \ \ B\psi_E^-=\sqrt E \psi_E^+,
\end{eqnarray}
which relates the eigenstates of SUSY partners which correspond the same non zero energy level.
It means that non-zero energy levels of SUSY Hamiltonian $H$ are two-fold degenerated and
eigenstates (\ref{psiup}), (\ref{psido}) correspond to the same energy level.
For SUSY Hamiltonian the lowest energy level is zero.
For zero energy level $E=0$ from (\ref{Eqzero}) we obtain the following first order differential equations
\begin{eqnarray}
B^+\psi_0^+={1\over\sqrt2}\left(-{d\over dx}+W(x)\right)\psi_0^+=0, \\
B\psi_0^-={1\over\sqrt2}\left({d\over dx}+W(x)\right)\psi_0^-=0
\end{eqnarray}
solutions of which are
\begin{eqnarray}
\psi_0^+=c_+\exp\left(\int_0^xdxW(x)\right), \\
\psi_0^-=c_-\exp\left(-\int_0^xdxW(x)\right).
\end{eqnarray}
From this it is clear that only one of the solutions can be square integrable.
We choose the superpotential that satisfies the following condition
\begin{eqnarray}
{\rm sign} W(x)=\pm 1, \ \ {\rm at} \ \ x\to\pm \infty.
\end{eqnarray}
Then  square integrable solution exists only for $\psi_0^-$ and thus zero energy ground state
of SUSY Hamiltonian is non-degenerated
\begin{eqnarray}\label{psi0-}
\psi_0=\psi_0^-(x)|\downarrow\rangle.
\end{eqnarray}
For review of SUSY quantum mechanics see, for instance, \cite{Coo95,Gan17}.

Now let us consider the question concerning the entanglement.
The zero energy ground state (\ref{psi0-}) is the product of coordinate wave function $\psi_0^-(x)$ and spin state $|\downarrow\rangle$.
We can say that the coordinate (continuous variable) and spin variable (discrete  variable) are separated.
So, entanglement of continuous variable and discrete spin variable in zero energy ground state
of SUSY Hamiltonian is zero.

Non-zero energy levels are two-fold degenerated and in general their eigenstate is superposition of two
states (\ref{psiup}) and (\ref{psido})
\begin{eqnarray}\label{psiGen}
\psi_E=c_1\psi_E^+(x)|\uparrow\rangle+c_2\psi_E^-(x)|\downarrow\rangle,
\end{eqnarray}
where normalization condition gives $|c_1|^2+|c_2|^2=1$.
In general case when $c_1\ne 0$ and $c_2\ne 0$ spin variable is entangled with continuous variable.
In the next section we consider measure of entanglement in this case.

\section{Entanglement of SUSY quantum states}
There are many well known definitions
for quantifying the entanglement in pure state. Amount them are, for instance, entropy entanglement,
concurrence, geometric measure of entanglement. In paper \cite{Fry17} it was found that geometric measure of entanglement
of spin with other quantum system can be
expressed by mean value of spin. In fact it was shown that coefficients in Schmidt decomposition
\begin{eqnarray}
\psi=\lambda_1\phi_1|\alpha_1\rangle+\lambda_2\phi_2|\alpha_2\rangle,
\end{eqnarray}
where states $\phi_1$ and $\phi_2$ are orthogonal as well as $|\alpha_1\rangle$ and $|\alpha_1\rangle$
are orthogonal can be expressed by the
mean value of spin
\begin{eqnarray}\label{ls}
\lambda_{1,2}^2={1\over 2}(1\pm |\langle{\bm\sigma}\rangle|).
\end{eqnarray}
Here $|\langle{\bm \sigma}\rangle|=\sqrt{\langle{\bm\sigma}\rangle^2}$. This result means that in order to obtain
coefficients in Schmidt decomposition it is not necessary to find Schmidt decomposition explicitly.
It is enough to calculate the mean value of spin with a given initial state. Then Schmidt coefficients are entirely determined
by the mean value of spin according to (\ref{ls}).
Having Schmidt coefficient we can easily find concurrence
\begin{eqnarray}\label{Csigma}
C=2\lambda_1\lambda_2=\sqrt{1-\langle{\bm\sigma}\rangle^2}.
\end{eqnarray}
In relation with this result it is worth to mention that
the geometric measure of entanglement as function of mean value of spin was obtained in \cite{Fry17}.

Note that relation of entanglement with mean value of spin is important also from the experimental point of view.
In order to measure entanglement of spin with other quantum system it is necessary to measure the mean value of spin.
For $|\langle{\bm\sigma}\rangle|=0$ the spin is maximally entangled with other quantum system.
Increasing of $|\langle{\bm\sigma}\rangle|$ leads to decreasing of entanglement. For $|\langle{\bm\sigma}\rangle|=1$
(it is maximally possible value in the case of pure state) the spin is disentangled.

Now let us find measure of entanglement of spin with continuous variable in
state (\ref{psiGen}). For this purpose we calculate the mean value of spin in this state
\begin{eqnarray}\nonumber
\langle\sigma_x\rangle=c_1^*c_2\langle\psi_E^+|\psi_E^-\rangle+c_1c_2^*\langle\psi_E^-|\psi_E^+\rangle= \\
=2|c_1||c_2||\langle\psi_E^+|\psi_E^-\rangle|\cos(\alpha_2-\alpha_1+\beta), \\  \nonumber
\langle\sigma_y\rangle=-ic_1^*c_2\langle\psi_E^+|\psi_E^-\rangle+ic_1c_2^*\langle\psi_E^-|\psi_E^+\rangle= \\
=2|c_1||c_2||\langle\psi_E^+|\psi_E^-\rangle|\sin(\alpha_2-\alpha_1+\beta),\\
\langle\sigma_z\rangle=|c_1|^2-|c_2|^2,
\end{eqnarray}
where
\begin{eqnarray}
c_1=|c_1|e^{i\alpha_1},\ \ c_2=|c_2|e^{i\alpha_2}, \ \ \langle\psi_E^+|\psi_E^-\rangle=|\langle\psi_E^+|\psi_E^-\rangle|e^{i\beta}.
\end{eqnarray}
Then we obtain
\begin{eqnarray}
\langle\bm{\sigma}\rangle^2=1-4|c_1|^2|c_2|^2(1-|\langle\psi_E^+|\psi_E^-\rangle|^2),
\end{eqnarray}
here we use that normalization condition $|c_1|^2+|c_2|^2=1$ gives $(|c_1|^2-|c_2|^2)^2=1-(|c_1|^2+|c_2|^2)^2+(|c_1|^2-|c_2|^2)^2=1-4|c_1|^2|c_2|^2$.
As result the concurrence reads
\begin{eqnarray}\label{Csusy}
C=2|c_1||c_2|\sqrt{(1-|\langle\psi_E^+|\psi_E^-\rangle|^2)}.
\end{eqnarray}
Entanglement is maximal if $|c_1|=|c_2|=1/\sqrt2$. It reads
\begin{eqnarray}\label{Cmax}
C_{max}=\sqrt{(1-|\langle\psi_E^+|\psi_E^-\rangle|^2)}.
\end{eqnarray}
If in addition $\psi_E^+(x)$ and $\psi_E^-(x)$ are orthogonal, their scalar product is zero $\langle\psi_E^+|\psi_E^-\rangle=0$.
Therefore  concurrence achieve its possible maximal value $C=1$. It takes place in the case of odd superpotential $W(-x)=-W(x)$ (when potential energy is
even function with respect to inversion of coordinate).

It is interesting to note that eigenstates of supercharges (\ref{Superchar}) are maximally entangled.
One can find that generators of SUSY act on states (\ref{psiup}), (\ref{psido}) as follows
\begin{eqnarray}
Q^+\psi_{\uparrow}=0, \ \  Q^+\psi_{\downarrow}=\sqrt E \psi_{\uparrow},\\
Q^-\psi_{\uparrow}=\sqrt E\psi_{\downarrow}, \ \ Q^-\psi_{\downarrow}=0.
\end{eqnarray}
Using this results we can easily solve the eigenvalue equation
$Q_1\psi_q^{(1)}=q\psi_q^{(1)}$
for supercharge $Q_1=Q^++Q^-$. We find eigenvalue $q=\pm\sqrt E$ with
eigenstates
\begin{eqnarray}
\psi_{\pm\sqrt E}^{(1)}={1\over\sqrt 2}(\psi_{\uparrow}\pm\psi_{\downarrow}).
\end{eqnarray}
Similarly eigenvalues of $Q_2=(Q^+-Q^-)$ are $q=\pm\sqrt E$ with the corresponding eigenstates
\begin{eqnarray}
\psi_{\pm\sqrt E}^{(2)}={1\over\sqrt 2}(\psi_{\uparrow}\pm  i\psi_{\downarrow}).
\end{eqnarray}
For this states $|c_1|=|c_2|=1/\sqrt2$ and spin is maximally entangled with continuous variable.
Thus eigenstates of supercharges are maximally entangled.
The case when the states are the eigenstates of supercharges look as abstract one.
In the next section we show that in fact this case can be realized in the nature.

\section{SUSY and entanglement in Jaynes-Cummings model}

The Hamiltonian of Jaynes-Cummings model
describes two level atom interacting with one electromagnetic mode.
In the resonance case, namely in the case when transition energy between two levels is equal to  energy of photon
the Hamiltonian reads
\begin{eqnarray}
H=H_0+H_{\rm int},
\end{eqnarray}
where
\begin{eqnarray}
H_0=\omega\left(b^+b+{1\over 2}\sigma_z\right)
\end{eqnarray}
describes noninteracting two level atom and photons, and
\begin{eqnarray}
H_{\rm int}=\gamma(b\sigma^++b^+\sigma^-)
\end{eqnarray}
describes the interaction of two-level atom with photons.

Hamiltonian $H_0$ is a simple example of SUSY system. We can write it in the form
\begin{eqnarray}
H_0=Q^2-{1\over 2},
\end{eqnarray}
where
\begin{eqnarray}
Q=b\sigma^++b^+\sigma^-.
\end{eqnarray}
One can see that there is Witten operator $\sigma_z$ anticommuting with supercharge $\{\sigma_z,Q\}=0$.
So, we have two supercharges $Q_1=Q$, $Q_2=i\sigma_zQ$ which together with $H_0$ satisfy $N=2$ SUSY algebra.
This algebra explains two-fold degeneration of exited energy levels of Hamiltonian $H_0$
\begin{eqnarray}
E_{n,\sigma}^0=\omega\left(n+{1\over 2}\sigma\right).
\end{eqnarray}
The eigenstates corresponding to these levels are
$|n\rangle|\sigma\rangle,$
where $\sigma=1$ corresponds to $|\uparrow\rangle$ and $\sigma=-1$ corresponds to $|\downarrow\rangle$.

Ground state energy is $E_{0,-1}=-\omega/2$ with state $|0\rangle|\downarrow\rangle$.
Excited states $|n-1\rangle|\uparrow\rangle$ and $|n\rangle|\downarrow\rangle$ have the same energies $E^0_{n-1,1}=E^0_{n,-1}=\omega(n-1/2)$.

The total Hamiltonian reads
\begin{eqnarray}\label{HQ2Q}
H=\omega Q^2 +\gamma Q-{1\over 2}.
\end{eqnarray}
Eigenstates for this Hamiltonian correspond to eigenstates for supercharge $Q$ which are maximally entangled.
Thus eigenstates of $H$ are maximally entangled too.
One can easily find solution of the eigenvalue equation for supercharge
\begin{eqnarray}
Q\psi_q=q\psi_q.
\end{eqnarray}
For $q=0$ we have eigenstate
\begin{eqnarray}
\psi_{q=0}=|0\rangle|\downarrow\rangle.
\end{eqnarray}
For $q=\pm\sqrt n$
we have  eigenstates
\begin{eqnarray}\label{JCpsi}
\psi_{q=\pm\sqrt n}={1\over\sqrt 2}(|n+1\rangle|\uparrow\rangle\pm |n\rangle|\downarrow\rangle).
\end{eqnarray}
These states are also eigenstates of the total Hamiltonian with eigenvalues
\begin{eqnarray}
E_{n,\pm}=\omega n\pm\gamma\sqrt n -{1\over 2}\omega.
\end{eqnarray}
For eigenstates (\ref{JCpsi}) spin is maximally entangled with continuous variable states.
In this case continuous variable states are orthogonal $\langle n+1|n\rangle=0$, therefore
according to (\ref{Cmax}) concurrence takes maximally possible value $C=1$.

\section{Conclusion}

In SUSY quantum mechanics nonzero energy levels  are two fold degenerated and the corresponding eigenstates in general are superposition of two states with spin up
and spin down (\ref{psiGen}). As result spin degree of freedom and continuous one in SUSY state are entangled.
We have calculated explicitly such measure of entanglement as concurrence for SUSY states and found that the concurrence is fully determined by mean value of spin in SUSY state (\ref{Csigma}) (final result for concurrence is presented by (\ref{Csusy})). We have concluded that concurrence achieves its maximal value for eigenstates of
supercharges. As an example  SUSY and entanglement in Jaynes-Cummings model which describes two levels atom interacting with one
electromagnetic mode  have been examined in resonance case when transition energy between two levels is equal to  energy of photon.
The total Hamiltonian of Jaynes-Cummings model can presented as linear combination of supercharge $Q$ and squared supercharge $Q^2$ (\ref{HQ2Q}).
Note that linear term $Q$ in Hamiltonian leads to splitting of energy levels (energy levels are not two fold degenerated).
In this case  the eigenstates of Jaynes-Cummings model corresponds to eigenstates of supercharge and entanglement of atom state with photon is maximal ($C=1$).

\section*{Acknowledgments}

This work was supported in part  by the project $\Phi\Phi$-83$\Phi$ (No. 0119U002203) from the Ministry of Education and Science of Ukraine.

\end{document}